%% file: Absolute-brs.tex
\documentclass[12pt,a4paper]{article}

\usepackage{ifthen} 
\newboolean{pdflatex}
\setboolean{pdflatex}{true} 
 
\newboolean{articletitles}
\setboolean{articletitles}{true} 
 
\newboolean{uprightparticles}
\setboolean{uprightparticles}{false} 
 
\newboolean{inbibliography}
\setboolean{inbibliography}{false} 

\usepackage{microtype}
\usepackage{lineno}  
\usepackage{xspace} 
 
\usepackage{graphicx}  
\usepackage{color}
\usepackage{colortbl}
 
\usepackage{amsmath} 
\usepackage{amssymb}
\usepackage{amsfonts}
\usepackage{upgreek} 
 
\newcommand*\patchAmsMathEnvironmentForLineno[1]{%
\expandafter\let\csname old#1\expandafter\endcsname\csname #1\endcsname
\expandafter\let\csname oldend#1\expandafter\endcsname\csname
end#1\endcsname
 \renewenvironment{#1}%
   {\linenomath\csname old#1\endcsname}%
   {\csname oldend#1\endcsname\endlinenomath}%
}
\newcommand*\patchBothAmsMathEnvironmentsForLineno[1]{%
  \patchAmsMathEnvironmentForLineno{#1}%
  \patchAmsMathEnvironmentForLineno{#1*}%
}
\AtBeginDocument{%
\patchBothAmsMathEnvironmentsForLineno{equation}%
\patchBothAmsMathEnvironmentsForLineno{align}%
\patchBothAmsMathEnvironmentsForLineno{flalign}%
\patchBothAmsMathEnvironmentsForLineno{alignat}%
\patchBothAmsMathEnvironmentsForLineno{gather}%
\patchBothAmsMathEnvironmentsForLineno{multline}%
}
 
\usepackage{hyperref}    
\usepackage[all]{hypcap} 
 
\input{lhcb-symbols-def} 
 
\usepackage{cite} 
\usepackage{mciteplus}
\begin{document}

\begin{titlepage}

\vspace*{-1.5cm}
\vspace*{1.5cm}
\hspace*{-5mm}\begin{tabular*}{16cm}{lc@{\extracolsep{\fill}}r}
 & &SYR-HEP-2019-001 \\ 
 & & \today \\ 
\end{tabular*}

\vspace*{4.0cm}

{\bf\boldmath\huge
\begin{center}
Methods of determining the absolute branching fractions of \Lc baryons and \Ds mesons at hadron colliders

\end{center}
}

\vspace*{2.0cm}

\begin{center}
Sheldon Stone 
\bigskip\\
{\it\footnotesize
Physics Department
Syracuse University, Syracuse, NY, USA 13244-1130\\
}
\end{center}

\vspace{\fill}

\begin{abstract}
 \noindent
 Current absolute branching fractions of \Lc and \Ds mesons are limiting the determination of important physics observables. These branching fractions have only been measured at $e^+e^-$ colliders. Here techniques for measuring them at hadron colliders are discussed. The precision determination of $\Lc\to p K^-\pi^+$ seems possible with current data, however the corresponding determination of $D_s^+\to K^+K^-\pi^+$ will require more integrated luminosity.
\end{abstract}

\vspace*{2.0cm}


\vspace{\fill}

\end{titlepage}

\pagestyle{empty}  



\setcounter{page}{2}
\mbox{~}


\pagestyle{plain} 
\setcounter{page}{1}
\pagenumbering{arabic}


\section{Introduction}

Accurate absolute branching fractions of \Lc and \Ds mesons are needed to provide precise determinations of physical quantities involving \Lb and \Bs mesons. Examples are measurements of $|V_{cb}|$, $|V_{ub}|$ and ${\cal{B}}(\Bs\to\mu^+\mu^-)$. Thus far all absolute branching fractions of charmed hadrons have been measured at $e^+e^-$ colliders~\cite{PDG}.

\section{\boldmath The determination of the absolute \Lc branching fraction}

The $\Lc\to pK^-\pi^+$ decay mode provides the largest sample of events for branching fraction measurements due to its relatively large value and comparably large detection efficiency with respect to other modes. 
The current directly measured branching fractions are shown in Table~\ref{tab:Lc-abs}, along with their weighted average.\footnote{Note that the weighted average value accidentally is exactly the same as the PDG fit~\cite{PDG}.} The current uncertainty is 5.3\%, larger than required for precision measurements.
\begin{table}[hbt]
\begin{center}
\caption{Absolute measurements of ${\cal{B}}(\Lc\to p K^-\pi^+)$.  The first uncertainty is statistical and the second systematic. The average is computed adding these in quadrature and assuming that the systematic uncertainties are uncorrelated. \label{tab:Lc-abs}}
  \begin{tabular}{lc}
\hline\hline  
Exp. &  ${\cal{B}}(\Lc\to p K^-\pi^+)$ (\%) \\\hline
Belle~\cite{Zupanc:2013iki}   & $6.84\pm 0.24^{+0.21}_{-0.27}$\\
BESIII~\cite{Ablikim:2015flg}  & ~~~$5.84\pm 0.27\pm0.23$\\
Wgt. Avg.   & $6.23\pm 0.33$
\\\hline
\hline
\end{tabular}
\end{center}
\end{table}

One solution to the problem of how to perform an absolute branching fraction measurement at hadron colliders is to find a $b$-hadron decay that allows determining the yield including all \Lc decays, and also finding the same decay into a specific \Lc final state, such as $pK^-\pi^+$, though any mode can be used. Methods to reconstruct $b$-hadron decays with a single missing massless particle have been used for semileptonic decays, e.g. $\Lb\to\Lc\mu^-\overline{\nu}$~\cite{Aaij:2017svr}, and are updated below for a missing heavy particle. 


Specifically, let us consider a $\overline{B}$ meson decaying into a $\PSigma_c(2455)$, a $\overline{p}$ and  possibly additional charged pions,  with a subsequent strong decay of the $\PSigma_c(2455)$ into $\pi^{\pm}\Lc$. Promising modes with their measured branching fractions are listed in  Table~\ref{tab:Bdecays}.\footnote{Mention of a decay mode implies the use of the charge-conjugate mode as well.} Note that the $\PSigma_c(2455)^0$ mass of $2453.75\pm 0.14$~\! MeV is nearly equal to  the $\PSigma_c(2455)^{++}$ mass of $2453.97\pm 0.14$~\! MeV. We also list related decays of the \Lb baryon.  As far as the signal yields are concerned, the \Lb decay branching fractions are of the same order as that of the $\overline{B}$ mesons, but the \Lb production fraction is substantially lower. There are many potential background channels. In the case of the $\overline{B}$ meson decays, the presence of the proton limits these.

\begin{table}[hbt]
\begin{center}
\caption{Measurements of useful  ${\cal{B}}(B \to \PSigma_c(2455) X)$ decays ~\cite{PDG}. \label{tab:Bdecays}}
  \begin{tabular}{lc}
\hline\hline  
Decay & ${\cal{B}}$  \\\hline
$B^-\to \PSigma_c(2455)^0\overline{p}$ &$(3.0\pm0.7)\cdot 10^{-5}$\\
$B^-\to \PSigma_c(2455)^0\overline{p}\pi^+\pi^-$ &$(3.5\pm 1.1)\cdot 10^{-4}$\\
$B^-\to \PSigma_c(2455)^{++}\overline{p}\pi^-\pi^-$ &$(2.4\pm 0.2)\cdot 10^{-4}$\\
$\Bzb\to \PSigma_c(2455)^{0}\overline{p}\pi^-$ &$(1.1\pm 0.2)\cdot 10^{-4}$\\
$\Bzb\to \PSigma_c(2455)^{++}\overline{p}\pi^+$ &$(1.9\pm 0.2)\cdot 10^{-4}$\\
$\Lb\to\PSigma_c(2455)^{0}\pi^+\pi^-$ &$(5.7\pm 2.2)\cdot 10^{-4}$\\
$\Lb\to\PSigma_c(2455)^{++}\pi^-\pi^-$ &$(3.2\pm 1.6)\cdot 10^{-4}$
\\\hline
\hline
\end{tabular}
\end{center}
\end{table}

\begin{figure}[b]
 \centering
\includegraphics[width=.95\textwidth]{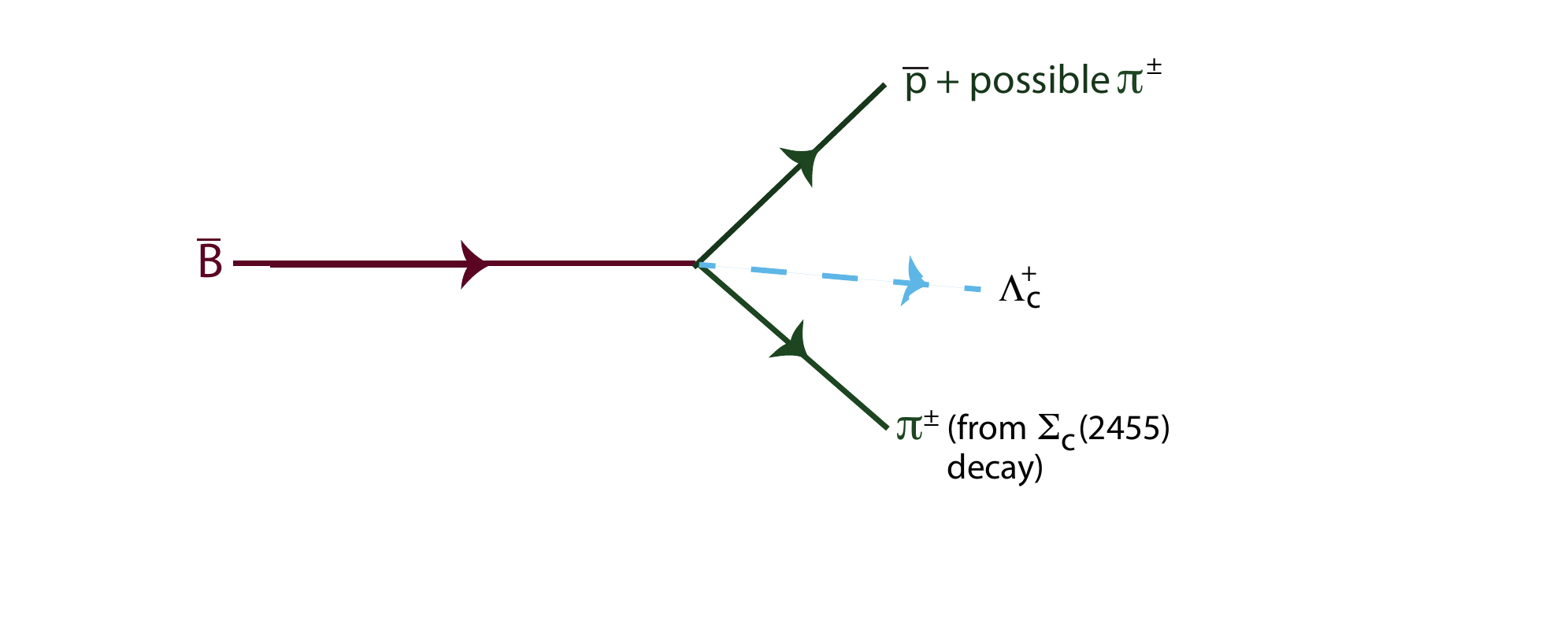}
\vspace{-1.5cm}
\caption{Topological illustration of a $\overline{B}$ meson decay into a $\overline{p}$, with possible additional charged pions, and a $\PSigma_c(2455)$ baryon that decays into $\pi^{\mp}\Lc$.}
 \label{geometry}
\end{figure}

The topology for the decays under consideration are shown in Fig.~\ref{geometry}.
Let us first consider the case where we do not attempt to find the \Lc decay. Taking the $\overline{B}$ direction as the $z$-axis, and having a $\overline{p}$ plus at least one additional charged pion not from the $\PSigma_c(2455)$ decay, the energy and momentum conservation equations can be written as assuming only one pion and a pion from the $\PSigma(2455)$ decay ($\pi_{\PSigma})$:
\begin{align}
\label{eq:G4C}
E_B~ &=E_{{p}}+ E_{\pi}+E_{\pi_{\PSigma}}+E_{\Lc},\\\nonumber 
p_{Bx} &=0=p_{p x}+p_{\pi x}+p_{\pi_{\PSigma} x}+p_{\Lc x},\\\nonumber
p_{By} &=0=p_{p y}+p_{\pi y}+p_{\pi_{\PSigma} y}+p_{\Lc y},\\\nonumber
p_{Bz} &=~~~~~p_{p z}+p_{\pi z}+p_{\pi_{\PSigma} z}+p_{\Lc z}.\\\nonumber
\end{align}

There are four unknowns $E_B$ (or $p_{Bz}$), and the three momenta of the missing \Lc, if we assume that the missing momenta is to be paired with the mass of the \Lc. The solution of these equations is derived in Appendix A, resulting in 
\begin{equation}
p_{B z}=\frac{4m_H^2p_{Az}\pm\sqrt{\left(4m_H^2p_{Az}\right)^2-16\left(E_A^2-p_{A z}^2\right)\left(4m_B^2E_A^2-m_H^4\right)}}{8\left(E_A^2-p_{A z}^2\right)}.
\end{equation}
where $E_A=E_p+E_{\pi}+E_{\pi \PSigma}$, $\overrightarrow{p}_{\!\!A}=\overrightarrow{p}_{\!\!p} +\overrightarrow{p}_{\!\!\pi}+\overrightarrow{p}_{\!\!\pi_{\PSigma}}$, and $m_H^2=-m_B^2-m_{A }^2+m_{\Lc}^2$.\footnote{It is easy to modify the equations for larger numbers of pions by adding them to the expressions for $E_A$ and $\overrightarrow{p}_{\!\!A}$.}
There is a two-fold ambiguity.
This technique can be applied directly to the decay without a $\PSigma_c$ intermediate state, e.g. $\Bm\to \overline{p} \pi^{-}\pi^-\pi^+ \Lc$, whose branching fraction is $(2.2\pm 0.7)\times 10^{-3}$, or $\Bzb\to \overline{p} \pi^{-}\pi^+ \Lc$ with a branching fraction of $(1.0\pm 0.1)\times 10^{-3}$~\cite{PDG}. However the two fold ambiguity could be a serious issue in performing a precision measurement.\footnote{This ambiguity could be suppressed by use of a $\overline{B}_{s2}^{*0}\to K^+B^-$ constraint as described in~\cite{Stone:2014mza} but with a loss of several hundred in statistics. This method has already been used in semileptonic \Bm decays~\cite{Aaij:2018unp}.} 

Another, superior method uses decays into $\PSigma_c(2455)$ baryons which provides an additional constraint that the  $\pi\Lc$ mass must equal the $\PSigma_c(2455)$ mass. (A better implementation is to constrain the mass difference between the unseen \Lc plus pion minus the  \Lc mass derived from the constraint equations to be equal to the known mass difference of the $\PSigma_c(2455)$ baryon with the \Lc.) The $\PSigma_c^0$ has a relatively narrow width of $1.83^{+0.11}_{-0.19}$~MeV, while the  $\PSigma_c^{++}$ has a similar width of $1.89^{+0.09}_{-0.18}$~MeV, both very favorable for applying a constraint. The additional charged pion from the $\PSigma_c$ decay also can help the precision of the $\overline{B}$ decay vertex. The $\overline{B}$ flight distance and its error are the most important factors that determine the $\overline{B}$ momentum resolution, and therefore the resolution on the quantities of interest.
The constraint equation is given in Appendix B.

The statistical limitation on the branching fraction measurement will be the smaller sample of fully reconstructed events.
LHCb measured a yield of $25\pm6$ events in the sum of the modes $\Lb\to\PSigma_c(2455)^{++}\pi^-\pi^-$ and $\PSigma_c(2455)^{0}\pi^+\pi^-$ using 35\invpb of 7\tev $pp$ collision luminosity~\cite{Aaij:2011rj}. Analyzing the full Run-I sample of 3\invfb should yield 2140. The branching fractions for these modes sum to $8.9\times 10^{-4}$, while the five $\overline{B}$ meson modes also shown in Table~\ref{tab:Bdecays} sum to $10^{-3}$. Since the event topologies are similar, we can extrapolate  the yields to the current LHCb data sample corresponding to 9\invfb.  In Run-2 the integrated luminosity was 6\invfb, the $b$ cross-section is doubled, and the ratio of \Lb production to the sum of the \Bz or the \Bp is about 1/4. (Also, the LHCb trigger has been improved since the initial data taking, but this has not been included.) Therefore, we predict $\sim$20~\!000 signal events in the sum of all five of the $\overline{B}$ meson decay modes, yielding a possible statistical uncertainty on the $\Lc\to p K^-\pi^+$ branching fraction of about 0.7\%, if the background is small in the fully reconstructed modes, which is already indicated by the LHCb measurement~\cite{Aaij:2011rj}.

The largest systematic uncertainties in the fully reconstructed sample come from only the \Lc decay tracks, since the other tracks are the same in the inclusive modes. Tracking and particle identification efficiencies are each at the level of 0.7\% per track. These can be added in quadrature and then summed over the three tracks, giving about a 2\% systematic uncertainty. This is much better than the current 5.3\% uncertainty. There will be additional small errors due to the relative event selection of different \Lc decay modes, the relative trigger efficiencies, the signal fitting technique, etc.. The analysis will need to be carefully thought through to minimize these uncertainties. It is also possible that studies with greater statistics will allow the tracking and particle identification uncertainties to be lowered. 

Key then to the usefulness of this technique are the backgrounds in the missing \Lc sample which are subject to potentially large systematic uncertainties.  
A survey of measured $b$ decay branching fractions shows that requiring a well identified proton vastly minimizes the background sources. Decays that could contribute are mixtures of particles from the two $b$-hadrons in the event, charmless $\overline{B}$ or \Lb decays into a proton and anti-nucleon with missing neutral particles.  No measurements of these decays are available.

For the fully reconstructed sample there are standard techniques to reduce backgrounds to a minimal level while still maintaining good efficiency. Here the candidate reconstructed $\overline{B}$ meson mass is plotted after a fit to the decay vertex and pointing to the primary vertex is applied~\cite{Hulsbergen:2005pu}. To distinguish signal from background in the inclusive sample a similar fit can be done with the addition of a $\PSigma_c$ mass constraint (see Eqs.~\ref{eq:mdiff}, \ref{eq:mLcpi}, and \ref{eq:mLc}), and then view the missing mass squared which for signal peaks at $m_{\Lc}^2$
\begin{equation}
MM^2=(E_B-E_A)^2-(P_{Bz}-\overrightarrow{P_A})^2=m_B^2+m_A^2-2(E_AE_B-P_{B z}P_{A z}).
\end{equation}
We expect only one solution will satisfy the $\PSigma_c$ mass constraint, and the resolution on  $MM^2$ will be improved.
 It will depend strongly on the flight distance of the $\overline{B}$ meson, since the $\overline{B}$ decay vertex resolution is mostly independent of the flight distance. Therefore, the flight distance requirement will have to be chosen wisely as a balance between the statistical and systematic uncertainties. 

\section{\boldmath Measuring the $D_s^+\to K^+K^-\pi^+$ branching fraction}
Here we try and apply the above techniques to find a method for measuring the  $D_s^+\to K^+K^-\pi^+$ branching fraction.  This is made much more difficult because the decays of the $D_s^{*+}$ meson are via a photon or a $\pi^0$ and are difficult to detect with high efficiency and good resolution in current detectors at hadron colliders. According the the PDG~\cite{PDG}, ${\cal{B}}(D_s^+\to K^+K^-\pi^+)=(5.44\pm 0.18)$\% combining CLEO~\cite{Onyisi:2013bjt}, Belle~\cite{Zupanc:2013byn}, and BaBar~\cite{delAmoSanchez:2010jg} measurements, corresponding to an accuracy of 3.3\%.

Decays that can be used along with the above mentioned techniques, with potentially large branching fractions, are $\Bsb\to \D_s(2536)^+ \pi^-$, or $\D_s(2536)^+ \pi^-\pi^+\pi^-$, with $\D_s(2536)^+\to\pi^+\pi^-\Ds$ Only the single pion \Bsb mode has been measured with a product branching fraction of $(2.5\pm 0.8)\times 10^{-5}$ \cite{Aaij:2012mra}. Twenty events are observed in a 1\invfb sample of LHCb data collected in 7\tev $pp$ collisions. This scales to 300 events for all the Run-1 plus Run-2 data. Clearly not statistically powerful enough to make a precision measurement. Perhaps one should expect an equal or larger event yield from the three pion mode. This measurement will be possible in the LHCb Upgrade I, where 50\invfb will be collected and the hadronic trigger efficiency will almost double. It is also possible to add other \Ds decay modes., but these will each have smaller yields and larger backgrounds. Hopefully, BESIII or BelleII can determine this quantity better before then.

\section{Conclusions}
In conclusion, several possibilities are discussed for measuring an absolute branching fraction of the \Lc baryon at a hadron collider. The basic idea is to reconstruct a $b$-hadron decay both with and without reconstructing the final state \Lc. The most promising method uses $\overline{B}\to \PSigma_c(2455) \overline{p} n\pi^{\pm}$ transitions with the $\PSigma_c(2455)\to\pi\Lc$ decay as a constraint. The current uncertainty could be significantly reduced using already collected data. This would be very useful for many measurements including those of the CKM elements $|V_{cb}|$ and implicitly $|V_{ub}|$. Similar considerations for the absolute branching ratio of the $\Ds$ meson are less sanguine, but could be made using the $\D_s(2536)^+\to\pi^+\pi^-\Ds$ as a constraint. An absolute \Ds branching fraction measurement of good precision, however, will take much more data then currently available.

After this paper was submitted, we became aware of a similar approach using only the decay $B^-\to \PSigma_c(2455)^{++}\overline{p}\pi^-\pi^-$. In addition, for the case where the \Lc is not fully reconstructed, kinematic restrictions designed to reduce background from some channels were studied \cite{Contu:2014wya}. These authors were the first to consider using the $\PSigma_c^{++}$ decay to make the \Lc absolute branching fraction measurement.

I thank Marina Artuso and Steve Blusk for useful discussions. Support for this work was provided by the U. S. National Science Foundation.

\section*{Appendix A: Solution of Eq.~\ref{eq:G4C}}
For simplicity, we define $E_A=E_p+E_{\pi}+E_{\pi_{\PSigma}}$, $\overrightarrow{p}_{\!\!A}=\overrightarrow{p}_{\!\!p} +\overrightarrow{p}_{\!\!\pi}+\overrightarrow{p}_{\!\!\pi_{\PSigma}}$.
Then we have
\begin{align}
\label{eq:G4C2}
&E_B=E_A+\sqrt{p_{\Lc x}^2+p_{\Lc y}^2+p_{\Lc z}^2+m_{\Lc}^2},\\\nonumber
&\left(E_B-E_A\right)^2=p_{\Lc x}^2+p_{\Lc y}^2+p_{\Lc z}^2+m_{\Lc}^2,\\\nonumber
&p_{\Lc x}=-p_{A x},\\\nonumber
&p_{\Lc y}=-p_{A y},\\\nonumber
&p_{\Lc z}=p_{Bz}-p_{A z},\\\nonumber
&\left(E_B-E_A\right)^2=p_{A x}^2+p_{A y}^2+p_{\Lc z}^2+m_{\Lc}^2,\\\nonumber
&\left(E_B-E_A\right)^2=p_{A x}^2+p_{A y}^2+(p_{B z}-p_{A z})^2+m_{\Lc}^2,\\\nonumber
\end{align}
Since $E_B^2=p_{B z}^2+m_B^2$, because $p_{B x}=p_{B y}=0$ we can write
\begin{align}
&E_B^2-2E_BE_A+E_A^2=p_{A x}^2+p_{A y}^2+(p_{B z}-p_{A z})^2 +m_{\Lc}^2,\\\nonumber
&p_{B z}^2+m_B^2-2\sqrt{p_{B z}^2+m_B^2}E_A+E_A^2=p_{A x}^2+p_{A y}^2+(p_{B z}-p_{A z})^2+m_{\Lc}^2,\\\nonumber
&-\sqrt{p_{B z}^2+m_B^2}=\left[-p_{B z}^2-m_B^2-E_A^2+p_{A x}^2+p_{A y}^2+(p_{B z}-p_{A z})^2+m_{\Lc}^2\right]/2E_A,\\\nonumber
&-\sqrt{p_{B z}^2+m_B^2}=\left[-m_B^2-E_A^2+p_{A }^2+m_{\Lc}^2-2p_{B z}p_{A z}\right]/2E_A,\\\nonumber
&-\sqrt{p_{B z}^2+m_B^2}=\left[-m_B^2-m_{A }^2+m_{\Lc}^2-2p_{B z}p_{A z}\right]/2E_A,\\\nonumber
&-\sqrt{p_{B z}^2+m_B^2}=\left[m_H^2-2p_{B z}p_{A z}\right]/2E_A,~~m_H^2=-m_B^2-m_{A }^2+m_{\Lc}^2.\\\nonumber
\end{align}
Squaring gives 
\begin{align}
&p_{B z}^2+m_B^2=\left[m_H^4+4p_{B z}^2p_{A z}^2-4m_H^2p_{B z}p_{A z}  \right]/4E_A^2\\\nonumber
&4p_{B z}^2\left(E_A^2-p_{A z}^2\right) -p_{Bz}\left(4m_H^2p_{Az}\right) +4m_B^2E_A^2-m_H^4=0.  \\\nonumber
\end{align}
This is a standard quadratic equation. The solution is
\begin{equation}
p_{B z}=\frac{4m_H^2p_{Az}\pm\sqrt{\left(4m_H^2p_{Az}\right)^2-16\left(E_A^2-p_{A z}^2\right)\left(4m_B^2E_A^2-m_H^4\right)}}{8\left(E_A^2-p_{A z}^2\right)}.
\end{equation}

\section*{\boldmath Appendix B: $\PSigma_c(2455)$ constraint equation}
First of all we have
\begin{equation}
\label{eq:mdiff}
m_{\Lc\pi_{\PSigma}}-m_{\Lc}=m_{\PSigma_c{(2455})}-m_{\Lc}=167.510\pm 0.017(\PSigma_c^{++}), ~167.290\pm 0.017(\PSigma_c^0).
\end{equation}
We next construct the $\Lc\pi_{\PSigma}$ invariant mass.
Once $p_{B z}$ is solved for, $p_{\Lc z}$ and  $E_{\Lc}$ can be determined. Furthermore
\begin{align}
&m_{\Lc\pi_{\PSigma}}^2=\left(E_{\Lc}+E_{\pi_{\PSigma}}\right)^2-\left(\overrightarrow{p}_{\!\!\Lc}+\overrightarrow{p}_{\!\!\pi_{\PSigma}}\right)^2\\\nonumber
&m_{\Lc\pi_{\PSigma}}^2=m_{\Lc}^2+m_{\pi_{\PSigma}}^2+2E_{\Lc}E_{\pi_{\PSigma}}-2\overrightarrow{p}_{\!\!\Lc}\cdot \overrightarrow{p}_{\!\!\pi_{\PSigma}}
\end{align}
Here $\overrightarrow{p}_{\!\!\Lc}=p_{\Lc x}\hat{i}+p_{\Lc y}\hat{j}+p_{\Lc z}\hat{k}$, where $\hat{i},~\hat{j},~{\rm and}~\hat{k}$ are unit vectors in the $x,~y,{\rm and}~z$ directions, respectively. Therefore
\begin{align}
&\overrightarrow{p}_{\!\!\Lc}=-p_{A x}\hat{i}-p_{A y}\hat{j}+(p_{B z}-p_{A z})\hat{k},\\\nonumber
&\overrightarrow{p}_{\!\!\pi_{\PSigma}}=p_{\pi_{\PSigma} x}\hat{i}+p_{\pi_{\PSigma} y}\hat{j}+p_{\pi_{\PSigma} z}\hat{k},\\\nonumber
\end{align}
where $A$ refers to either the $\overline{p}$ or a $\overline{p}$ plus $\pi_{\PSigma}$ combination, and  $p_{\pi_{\PSigma}}$ refers to the pion from the $\PSigma_c$ decay. Putting this together
\begin{align}
&E_{\Lc}E_{\pi_{\PSigma}}=\sqrt{p_{A x}^2+p_{A y}^2+(p_{B z}-p_{A z})^2}\sqrt{p_{\pi_{\PSigma} x}^2+p_{\pi_{\PSigma} y}^2+p_{\pi_{\PSigma} z}^2}\\\nonumber
&\overrightarrow{p}_{\!\!\Lc}\cdot\overrightarrow{p}_{\!\!\pi_{\PSigma}}=-p_{A x}p_{\pi_{\PSigma} x}-p_{A y}p_{\pi_{\PSigma} y}+(p_{B z}-p_{A z})p_{\pi_{\PSigma} z}\\\nonumber
\end{align}
Therefore
\begin{align}
\label{eq:mLcpi}
m_{\Lc\pi_{\PSigma}}^2=&m_{\Lc}^2+m_{\pi_{\PSigma}}^2+2\sqrt{p_{A x}^2+p_{A y}^2+(p_{B z}-p_{A z})^2}\sqrt{p_{\pi_{\PSigma} x}^2+p_{\pi_{\PSigma} y}^2+p_{\pi_{\PSigma} z}^2}\\\nonumber
&-2\left(-p_{A x}p_{\pi_{\PSigma} x}-p_{A y}p_{\pi_{\PSigma} y}+(p_{B z}-p_{A z})p_{\pi_{\PSigma} z}\right).\\\nonumber
\end{align}
Now for $m_{\Lc}$
\begin{align}
\label{eq:mLc}
m_{\Lc}^2&=E_{\Lc}^2-p_{\Lc x}^2-p_{\Lc y}^2-p_{\Lc z}^2\\\nonumber
&=\left(\sqrt{p_{B z}^2+m_B^2}-E_A\right)^2-p_{A x}^2-p_{A y}^2-\left(p_{B z}-p_{Az}\right)^2\\\nonumber
&=m_B^2+m_A^2-2E_A\sqrt{p_{B z}^2+m_B^2}+2p_{B z}p_{Az}.\\\nonumber
\end{align}

\newpage

\input{Absolute-brs.bbl}

\end{document}

%% file: lhcb-symbols-def.tex
\usepackage{xspace} 
\usepackage{upgreek}

\def\MagUp {\mbox{\em Mag\kern -0.05em Up}\xspace}

\ifthenelse{\boolean{uprightparticles}}%
{

 \def\PDelta      {\ensuremath{\Delta}\xspace}                 
 \def\PXi         {\ensuremath{\Xi}\xspace}                 
 \def\PLambda     {\ensuremath{\Lambda}\xspace}                 
 \def\PSigma      {\ensuremath{\Sigma}\xspace}                 
 \def\POmega      {\ensuremath{\Omega}\xspace}                 
 \def\PUpsilon    {\ensuremath{\Upsilon}\xspace}

 \def\PB      {\ensuremath{\mathrm{B}}\xspace}                 
                  
 \def\PD      {\ensuremath{\mathrm{D}}\xspace}

 \def\PK      {\ensuremath{\mathrm{K}}\xspace}

 \def\Pb      {\ensuremath{\mathrm{b}}\xspace}                 
 \def\Pc      {\ensuremath{\mathrm{c}}\xspace}

 \def\Pi      {\ensuremath{\mathrm{i}}\xspace}

 \def\Ps      {\ensuremath{\mathrm{s}}\xspace}

 \def\thebaroffset{0.0em}
}
{

 \mathchardef\PDelta="7101
 \mathchardef\PXi="7104
 \mathchardef\PLambda="7103
 \mathchardef\PSigma="7106
 \mathchardef\POmega="710A
 \mathchardef\PUpsilon="7107
                  
 \def\PB      {\ensuremath{B}\xspace}                 
                  
 \def\PD      {\ensuremath{D}\xspace}

 \def\PK      {\ensuremath{K}\xspace}

 \def\Pb      {\ensuremath{b}\xspace}                 
 \def\Pc      {\ensuremath{c}\xspace}

 \def\Pi      {\ensuremath{i}\xspace}

 \def\Ps      {\ensuremath{s}\xspace}

 \def\thebaroffset{0.18em}
}
\newcommand{\offsetoverline}[2][\thebaroffset]{\kern #1\overline{\kern -#1 #2}}%

\makeatletter
\ifcase \@ptsize \relax
  \newcommand{\miniscule}{\@setfontsize\miniscule{4}{5}}
\or
  \newcommand{\miniscule}{\@setfontsize\miniscule{5}{6}}
\or
  \newcommand{\miniscule}{\@setfontsize\miniscule{5}{6}}
\fi
\makeatother

\DeclareRobustCommand{\optbar}[1]{\shortstack{{\miniscule (\rule[.5ex]{1.25em}{.18mm})}
  \\ [-.7ex] $#1$}}

\def\squark    {{\ensuremath{\Ps}}\xspace}

\def\cquark    {{\ensuremath{\Pc}}\xspace}

\def\bquark    {{\ensuremath{\Pb}}\xspace}

\def\KorKbar {\kern \thebaroffset\optbar{\kern -\thebaroffset \PK}{}\xspace}

\def\D       {{\ensuremath{\PD}}\xspace}

\def\DorDbar {\kern \thebaroffset\optbar{\kern -\thebaroffset \PD}\xspace}

\def\Ds      {{\ensuremath{\D^+_\squark}}\xspace}

\def\B       {{\ensuremath{\PB}}\xspace}
\def\Bbar    {{\ensuremath{\offsetoverline{\PB}}}\xspace}

\def\BorBbar {\kern \thebaroffset\optbar{\kern -\thebaroffset \PB}\xspace}
\def\Bz      {{\ensuremath{\B^0}}\xspace}
\def\Bzb     {{\ensuremath{\Bbar{}^0}}\xspace}
\def\Bu      {{\ensuremath{\B^+}}\xspace}
\def\Bub     {{\ensuremath{\B^-}}\xspace}
\def\Bp      {{\ensuremath{\Bu}}\xspace}
\def\Bm      {{\ensuremath{\Bub}}\xspace}

\def\Bs      {{\ensuremath{\B^0_\squark}}\xspace}
\def\Bsb     {{\ensuremath{\Bbar{}^0_\squark}}\xspace}

\def\BorBsbar {\kern \thebaroffset\optbar{\kern -\thebaroffset \PB_s}\xspace}

\def\Y#1S{\ensuremath{\PUpsilon{(#1S)}}\xspace}

\def\Lz          {{\ensuremath{\PLambda}}\xspace}

\def\LorLbar     {\kern \thebaroffset\optbar{\kern -\thebaroffset \PLambda}\xspace}

\def\Lc          {{\ensuremath{\Lz^+_\cquark}}\xspace}

\def\Lb           {{\ensuremath{\Lz^0_\bquark}}\xspace}

\def\to                 {\ensuremath{\rightarrow}\xspace}

\def\AT#1     {\ensuremath{A_{\mathrm{T}}^{#1}}\xspace}           

\def\C#1      {\ensuremath{\mathcal{C}_{#1}}\xspace}                       
\def\Cp#1     {\ensuremath{\mathcal{C}_{#1}^{'}}\xspace}                    
\def\Ceff#1   {\ensuremath{\mathcal{C}_{#1}^{\mathrm{(eff)}}}\xspace}        
\def\Cpeff#1  {\ensuremath{\mathcal{C}_{#1}^{'\mathrm{(eff)}}}\xspace}       
\def\Ope#1    {\ensuremath{\mathcal{O}_{#1}}\xspace}                       
\def\Opep#1   {\ensuremath{\mathcal{O}_{#1}^{'}}\xspace}                    

\newcommand{\aunit}[1]{\ensuremath{\text{\,#1}}}       

\newcommand{\tev}{\aunit{Te\kern -0.1em V}\xspace}
\newcommand{\gev}{\aunit{Ge\kern -0.1em V}\xspace}
\newcommand{\mev}{\aunit{Me\kern -0.1em V}\xspace}
\newcommand{\kev}{\aunit{ke\kern -0.1em V}\xspace}
\newcommand{\ev}{\aunit{e\kern -0.1em V}\xspace}
\newcommand{\mevc}{\ensuremath{\aunit{Me\kern -0.1em V\!/}c}\xspace}
\newcommand{\gevc}{\ensuremath{\aunit{Ge\kern -0.1em V\!/}c}\xspace}
\newcommand{\mevcc}{\ensuremath{\aunit{Me\kern -0.1em V\!/}c^2}\xspace}
\newcommand{\gevcc}{\ensuremath{\aunit{Ge\kern -0.1em V\!/}c^2}\xspace}

\def\pb {\aunit{pb}\xspace}
\def\invpb {\ensuremath{\pb^{-1}}\xspace}
\def\fb   {\ensuremath{\aunit{fb}}\xspace}
\def\invfb   {\ensuremath{\fb^{-1}}\xspace}

\def\gsim{{~\raise.15em\hbox{$>$}\kern-.85em
          \lower.35em\hbox{$\sim$}~}\xspace}
\def\lsim{{~\raise.15em\hbox{$<$}\kern-.85em
          \lower.35em\hbox{$\sim$}~}\xspace}

\xspace

\def\tell1  {TELL1\xspace}
\def\ukl1   {UKL1\xspace}



%% file: Absolute-brs.bbl
\ifx\mcitethebibliography\mciteundefinedmacro
\PackageError{LHCb.bst}{mciteplus.sty has not been loaded}
{This bibstyle requires the use of the mciteplus package.}\fi
\providecommand{\href}[2]{#2}